\documentclass[prl,preprintnumbers,twocolumn,floatfix,nofootinbib,a4paper,superscriptaddress]{revtex4-2}
\usepackage{color,amsmath}
\usepackage{graphicx}
\usepackage{dcolumn}
\usepackage{bm}
\usepackage{macros}

\begin{document}


\title{Suppressed Magnetogenesis from Ultralight Dark Matter due to Finite Conductivity}

\author{Ramkishor Sharma}
\email{ramkishor.sharma@uohyd.ac.in}
\author{Samarth Majumdar}%
\email{samarth.majumdar@gmail.com}
\affiliation{School of Physics, University of Hyderabad, Central University P.O, Hyderabad-500046, Telangana, India.}
\author{Divya Sachdeva}
\email{divyasachdeva@phy.iith.ac.in}
\affiliation{Indian Institute of Technology Hyderabad, Kandi, Sangareddy-502285, Telangana, India}

\date{\today}

\begin{abstract}
Recently, a mechanism for generating astrophysically relevant magnetic fields via ultralight pseudoscalar dark matter, through the coupling term $g_{\phi \gamma} \phi F_{\mu \nu}\tilde{F}^{\mu\nu}$ in the Lagrangian density, was proposed in Ref.~\cite{Brandenberger:2025gks}. In this scenario, the electromagnetic fields are amplified through the phenomena of parametric resonance due to the oscillatory behaviour of the pseudoscalar field. However, the analysis presented in that work does not account for the effects of a conducting medium. In this paper, we incorporate the finite conductivity of the plasma into the dynamics of the pseudoscalar and electromagnetic fields. We show that, due to the large conductivity relative to the Hubble parameter, the amplification of the electromagnetic fields due to parametric resonance is significantly suppressed. Consequently, we find that, for observationally viable values of the coupling between the electromagnetic field and the ultralight pseudoscalar field, it is not possible to generate magnetic fields of sufficient strength to explain their presence in cosmic voids.
\end{abstract}

\maketitle

\noindent\textbf{Introduction.---}
The presence of magnetic fields in collapsed structures such as galaxies and galaxy clusters has been well established through a wide range of observational studies \cite{Widrow:2002ud,Beck:2008ty,Bernet:2008qp,Kronberg:2007dy}. There is also indirect evidence for magnetic fields in regions far from such collapsed objects, namely in cosmic voids \cite{Neronov:1900zz, Finke:2015ona}. The origin of these magnetic fields has been extensively investigated in the literature. In collapsed structures, magnetic fields can be explained by the amplification, via the dynamo mechanism, of a small seed field generated by astrophysical processes operating during structure formation or reionization \cite{Kulsrud1997,Gnedin2000,Sethi:2004pe}. However, explaining the presence of magnetic fields in voids through such astrophysical processes is challenging, which has motivated the study of primordial magnetic fields (see \RRef{Durrer:2013pga, Subramanian:2015lua, Vachaspati:2020blt} for reviews).

Recently, Ref.~\cite{Brandenberger:2025gks, Brahma:2026fre} investigated the generation of magnetic fields after recombination via ultralight pseudoscalar
dark matter through the phenomenon of parametric resonance due to the oscillating behaviour of the dark matter field. The authors found that the produced magnetic field energy density can easily be comparable to that of dark matter for observationally viable values of the coupling between the electromagnetic (EM) field and the pseudoscalar field. Moreover, the resulting magnetic field strength could potentially account for their presence on large scales in the Universe\footnote{Magnetogenesis mechanisms involving pseudoscalar-gauge field interactions have also been studied during inflation and reheating\cite{Turner:1987bw, Garretson:1992vt,Anber:2006xt,Fujita:2015iga,Adshead:2016iae,Gorbar:2021zlr,Yanagihara:2023qvx,Brandenburg:2024awd,Sharma:2024nfu,Iarygina:2025ncl}.}.

However, the analysis in Ref.~\cite{Brandenberger:2025gks} did not account for the effects of the conducting medium. Although the Universe transitions from an ionised state to a predominantly neutral phase after recombination, the residual ionization fraction remains of order $10^{-4}$. This finite ionization can influence the evolution of EM fields through their interaction with charged particles via the Lorentz force. In this letter, we study the dynamics of the ultralight pseudoscalar and the EM field by incorporating the finite conductivity of the post-recombination medium. We find that, once this effect is included, the amplification of the EM field is significantly suppressed compared to the results of Ref.~\cite{Brandenberger:2025gks}, and is insufficient to generate magnetic fields of astrophysical interest.



\noindent\textbf{Magnetogenesis driven with a pseudo scalar dark matter post recombination.---}
The dynamics of the ultra light pseudo scalar dark matter and the EM field is described by the action;
\begin{align}\label{Action_axion_U1}
    S&=\int d^4x \sqrt{-g} \Bigg[-\frac{\partial_\mu\phi \partial^\mu\phi}{2}-V(\phi)-\frac{F_{\mu \nu}F^{\mu\nu}}{4}\nonumber\\
    &-\frac{g_{\phi \gamma} \phi}{4} F_{\mu \nu}\tilde{F}^{\mu\nu}\Bigg].
\end{align}
Here, $\phi$ and $V(\phi) \equiv m^2 \phi^2/2$ denotes the pseudoscalar dark matter and its potential respectively and $F_{\mu \nu} \equiv \partial_{\mu}A_{\nu}-\partial_{\nu} A_{\mu}$ denotes the EM field strength tensor. $g_{\phi \gamma}$ denotes the coupling strength in between the $\phi$ and EM field. Observationally, it has been constrained to be $8 \times 10^{-13} {\rm GeV^{-1}}$ for $m < 10^{-12} {\rm eV}$ \cite{Reynolds:2019uqt,Reynes:2021bpe,CAST:2024eil,Fedderke:2019ajk}. Assuming that the background evolution is described by a spatially flat Friedmann-Lemaitre-Robertson-Walker (FLRW) metric, we get the following equation of motion for the $\phi$ and the EM field;
\begin{align}
    \Ddot{\phi}+3 H \dot{\phi}-\ddel \phi+ \frac{d V}{d \phi}=\frac{g_{\phi \gamma}}{a^4}\,\EE\cdot\BB&\label{phi_eqn}\,,\\
    \ddot{\AA}+H \dot{\AA}-\frac{1}{a}\Del \dot{A_0}-\frac{1}{a^2}\ddel \AA+\frac{1}{a^2}\Del(\Del \cdot \AA)&\nonumber\\
    -g_{\phi \gamma}\left(\frac{\dot{\phi}\BB}{a}+\frac{1}{a^2}\Del \phi \times \EE\right)=0.&\label{Ai_eqn}
\end{align}
Here $\BB=\Del \times \AA,~ \EE=-a \dot{\AA}+\Del A_0$ and
$H$ is the Hubble parameter with $H=\dot{a}/a$ for scale factor $a(t)$. In the case of ultralight pseudo scalar dark matter, the $\phi$ field remains frozen due to the Hubble drag during the time when the value of the Hubble parameter, $H$ is greater than its mass, $m$. Once the mass of the field becomes greater than the $H$, the field starts oscillating around its minima. For the parameter space of interest, taken in Ref.\cite{Brandenberger:2025gks}, the $\phi$ field is in oscillating phase during and after recombination. We discussed the details about the initial condition in the next section. By assuming that we are working in a coupling regime where the backreaction of the EM field on the $\phi$ dynamics and also the inhomogeneity of the $\phi$ field can be neglected, we get the following evolution equations in the Weyl gauge ($A_0=0$);
\begin{align}
    \Ddot{\phi} + 3 H \Dot{\phi} + m^2 \phi = 0, \quad H^{2} = \frac{1}{6 M_{\rm pl}^2} \bigg(\Dot{\phi}^{2} + &m^2 \phi^2 \bigg) ,\label{inflaton}\\
    \Ddot{\mathcal{A}}_{\pm} + H \dot{\mathcal{A}}_{\pm} + \frac{k}{a} \bigg( \frac{k}{a} \mp g_{\phi \gamma} \dot{\phi} \bigg) \mathcal{A}_{\pm} &= 0 \label{A_equation}.
\end{align}
Here, $\mathcal{A}_{\pm}$ denotes the vector potential in Fourier space, and $\pm$ in the suffix represents the two helicity modes in circular polarisation basis. In the above equation, we neglect the conductivity of the plasma. After including that, the \Eq{A_equation} modifies to,
\begin{equation}\label{A_equation_with_plasma}
    \Ddot{\mathcal{A}}_{\pm} + (H +\sigma)\dot{\mathcal{A}}_{\pm} + \frac{k}{a} \bigg( \frac{k}{a} \mp g_{\phi \gamma} \dot{\phi} \bigg) \mathcal{A}_{\pm} = 0 .
\end{equation}
Here $\sigma$ denotes the conductivity of the plasma. The above equation is obtained by introducing the current term, $\JJ = \sigma (\EE + \vv \times \BB)$, into \Eq{Ai_eqn} (see Sec.~II of Ref.~\cite{Long:2015cza} for details). Since we are primarily interested in the effect of the pseudoscalar field on the evolution of the EM field, we neglect the term involving the fluid velocity, $\vv$. Furthermore, using $\EE = -\dot{a \AA}$, we arrive at \Eq{A_equation_with_plasma} in Fourier space. To estimate the conductivity after recombination, we follow the analysis given in section (2.6.5) of Ref.~\cite{Shukurov_Subramanian_2021}. The conductivity can be written as
\begin{align}\label{cond_expr}
    \sigma \sim \frac{n_e e^2 \tau_c}{m_e},
\end{align}
where $n_e$ and $m_e$ denote the number density and mass of the electron, respectively. The quantity $\tau_c$ represents the characteristic collision time over which an electron reaches its terminal drift velocity. After recombination, collisions of the electron with neutral particles dominate the contribution in \Eq{cond_expr}, and the collision time is given by
\begin{equation}
    \tau_c =\frac{1}{n_b \langle\sigma_{ne} v\rangle} ,
\end{equation}
where $n_n$ denotes the number density of neutral particles (mainly Hydrogen and Helium atoms), $\langle\sigma_{ne} v\rangle$ is the thermally averaged scattering cross section of the electrons with the neutral particles. Substituting the expression for $\tau_c$ into the conductivity, we obtain
\begin{align}\label{sigma}
    \sigma &\sim \frac{n_e e^2}{n_b \langle\sigma_{ne} v\rangle} = \frac{x_e e^2}{m_e \langle\sigma_{ne} v\rangle} \sim 0.043 {\rm eV} \frac{x_e}{10^{-4}} \left(\frac{T}{0.3 {\rm eV}}\right)^{-0.68}
\end{align}
Here, $x_e$ denotes the ratio of the number density of electrons to that of the neutral particles. In the last equality of the above equation, we take the collision rate of electron and neutral hydrogen, $\langle\sigma_{ne} v\rangle \approx 4.8 \times 10^{-9} (T/10^2 K)^{0.68} {\rm cm^3 s^{-1}}$ (see Eq. 2.41 in Ref.~\cite{2011piim.book.....D}) and have neglected the effects of the Helium atoms.
Furthermore, we compare the conductivity with the Hubble parameter. We can express the Hubble parameter post recombination in terms of temperature as follows,
\begin{align}
    H &\sim \frac{1}{\sqrt{3} M_{pl}}\sqrt{\frac{2 g_{eq}\pi^2}{30} T_{eq}^4\left(\frac{T}{T_{eq}}\right)^3}\\
    &= 5.9 \times 10^{-29} {\rm eV} \left(\frac{T}{0.3 {\rm eV}}\right)^{3/2} \left(\frac{T_{eq}}{1 {\rm eV}}\right)^{1/2}.
\end{align}
Here, $T_{eq}$ denotes the temperature at radiation-matter equality. Using this and \Eq{sigma}, we get,
\begin{align} \label{sigma_by_H}
    \frac{\sigma}{H} \sim 7.3 \times 10^{26} \frac{x_e}{10^{-4}}\left(\frac{T}{0.3 {\rm eV}}\right)^{-2.18}
\end{align}
Furthermore in this section, we provide the simple estimates for the amplification of the EM field. From \Eq{A_equation_with_plasma}, one can infer that there is an instability for the modes $k< a g \dot{\phi}$. Since the $\phi$ field oscillates with time, the instability is only active during a fraction of each oscillation period and this instability lead to the amplification of each helicity mode alternatively. In our simple estimates, we provide the growth rate using the amplitude of $\dot{\phi}$. This, of course overestimates the amplification; we will provide a more accurate analysis by solving \Eq{inflaton} numerically to obtain the evolution of $\phi(t)$ and $\Dot{\phi}(t)$ in the next section. Further, we will use the solution of $\dot{\phi}$ to obtain the dynamics of the vector potential by solving \Eq{A_equation_with_plasma}.

As given in \Eq{sigma_by_H}, the conductivity after recombination is much larger than the Hubble parameter. Using this, one can neglect the Hubble friction term in \Eq{A_equation_with_plasma} in comparison to the term having conductivity. Furthermore, as the conductivity is very large, the system will be in an overdamped regime, and one can neglect the $\Ddot{\mathcal{A}}_{\pm}$ in comparison to other terms. Including this, the above equation reduces to,
\begin{equation}
    \sigma m \frac{d \mathcal{A}_{\pm}}{d z} + \frac{k}{a} \bigg( \frac{k}{a} \mp g_{\phi \gamma} \dot{\phi} \bigg) \mathcal{A}_{\pm} = 0 .
\end{equation}
 Here, $z=m t$. For the modes, $k< a g \dot{\phi}$, there is an exponential growth of each helicity mode, the growth rate is given by,
\begin{align}
    \mu_k \sim \frac{k}{a}\frac{g_{\phi \gamma} \dot{\phi}}{\sigma m}.
\end{align}
As the kinetic and potential energy are of similar order during the oscillations of the $\phi$ field, the \Eq{inflaton} implies $\dot{\phi} \sim H M_{pl}$. Using this we get $\mu_k \sim g_{\phi \gamma} M_{pl} (k/ a H) (H/m) (H/\sigma)$ . The maximum amplification takes for the wavenumber, $k_c \sim a g_{\phi \gamma} \dot{\phi} \sim a g_{\phi \gamma} H M_{pl}$. Using this, we get, $\mu_k \sim f_{k_c} (g_{\phi \gamma} M_{pl})^2 (H/m) (H/\sigma)$ for the mode $k$ which is a fraction ($f_{k_c}$) of $k_c$. Using this, the exponent of the amplification in time $z$ is given by,
\begin{align}
    \mu_k z = f_{k_c} (g_{\phi \gamma} M_{pl})^2 \frac{H}{m} \frac{H}{\sigma} m t.
\end{align}
The above quantity will take the following value in a time duration of one Hubble time ($t \sim 1/H$),
\begin{align}\label{ampl_cond}
    \mu_k z = f_{k_c} (g_{\phi \gamma} M_{pl})^2 \frac{H}{\sigma} = 10^{-13} f_{k_c}\left(\frac{g_{\phi \gamma} M_{pl}}{10^6}\right)^2 \frac{H/\sigma}{10^{-25}}.
\end{align}
As $e^{10^{-13}} \approx 1$, the above expression indicates that there is no amplification of the vector potential for $g_{\phi \gamma}\sim 10^{-12} {\rm GeV^{-1}}$ and $H/\sigma =10^{-25}$. As the observations typically provides an upper limit on $g_{\phi\gamma}$ of the order of $10^{-12} {\rm GeV^{-1}}$ for $m < 10^{-12} {\rm eV}$ \cite{Reynolds:2019uqt}, hence we have considered $g_{\phi\gamma} \sim 10^{-12} {\rm GeV^{-1}}$ in the estimates provided in \Eq{ampl_cond}. By demanding $\mu_k > H$, one can obtain a bound on the coupling strength, $g_{\phi \gamma}$ to have a significant amplification, This condition implies,
\begin{equation}
    g_{\phi \gamma} > \sqrt{\frac{\sigma}{H}} \frac{1}{M_{pl}} \sim 1.3 \times 10^{-6} {\rm GeV^{-1}} \sqrt{\frac{\sigma/H}{10^{25}}}.
\end{equation}
As this value of $g_{\phi\gamma}$ is observationally ruled out for $m< 10^{-12} {\rm eV}$, we conclude that, including the effects of the charged plasma, the magnetogenesis mechanism proposed in Ref~\cite{Brandenberger:2025gks} is not viable to produce the magnetic field of astrophysical interest. Here, we would also like to mention that we have derived the above result neglecting the backreaction of the EM field on the dynamics of the $\phi$ field. Since there is almost no amplification of the EM field, our assumption will hold.
However, if we do not include the conductivity in \Eq{A_equation_with_plasma}, the system will be in the weak damping regime as $H^2 << g_{\phi \gamma} \dot{\phi} (k/a)$ for $g_{\phi \gamma}< 10^{-12} {\rm GeV^{-1}}$ and $k \leq k_c$. In this case, $\mu_k z$ for $k=f_{k_c} k_c$ for $t\sim 1/H$ is given by,
\begin{align}
    \mu_k z \sim \sqrt{f_{k_c}\left(\frac{g_{\phi \gamma} M_{pl} H}{m}\right)^2} z\sim \times 10^{8} \sqrt{f_{k_c}}\frac{g_{\phi \gamma}}{10^{-10} {\rm GeV^{-1}}}.
\end{align}
This result is similar to equation (15) of Ref.~\cite{Brandenberger:2025gks}. 

\begin{figure}[h!]
\centering
 \includegraphics[width=0.5\textwidth]{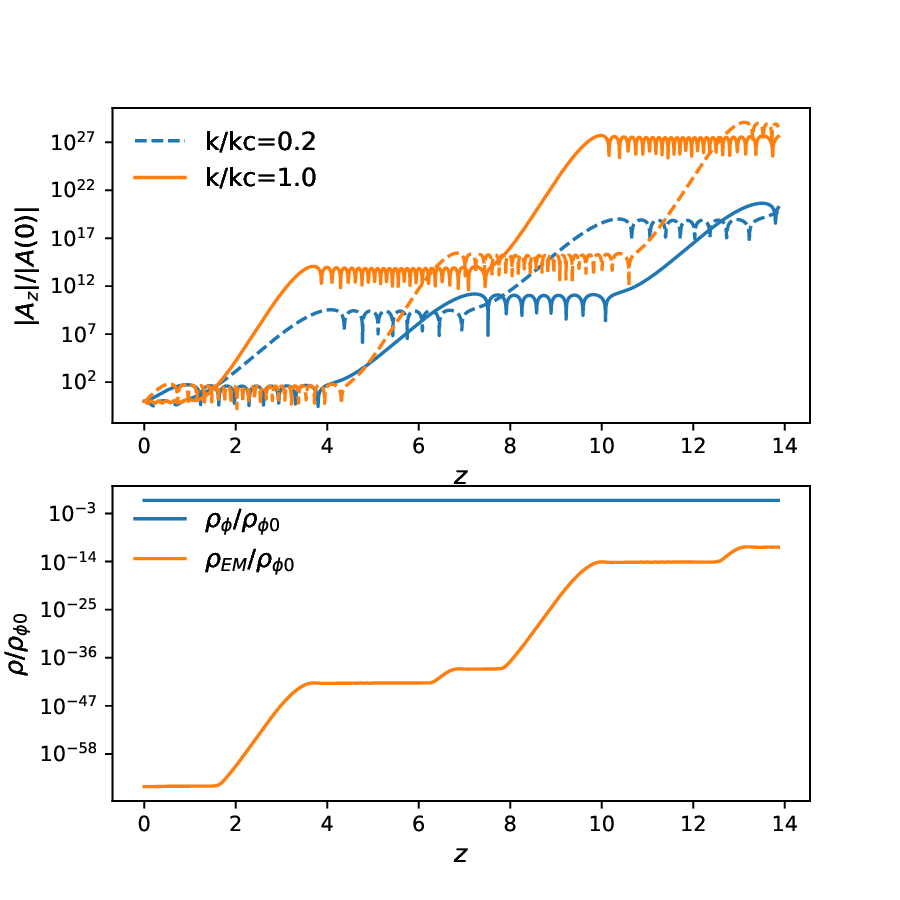}
\caption{The upper panel shows the time evolution of the two helicity modes of the vector potential, normalized to their initial values. The solid and dashed lines correspond to the `$-$' and `$+$' helicity modes, respectively. The results are shown for $k/k_c = 0.2$ and $1.0$. The lower panel displays the evolution of the energy densities of the $\phi$ field ($\rho_{\phi}$) and the EM field ($\rho_B$), both normalized to $\rho_{\phi}$ at $z=0$. In this figure, we have taken $\sigma=0$.}
\label{figure1}
\end{figure}

\noindent \textbf{Numerical estimates for the amplification of the EM field.---}
To accurately capture the dynamics of the $\phi$ field and the EM field, we solve \Eq{inflaton} numerically\footnote{The simulations in this study were performed using the publicly available Pencil Code \cite{PencilCode:2020eyn,PC}.}. Using the resulting $\dot{\phi}$, we then solve \Eq{A_equation_with_plasma} to obtain the vector potential. In the first case we neglect the conductivity in \Eq{A_equation_with_plasma} and we choose $m=10^{-18} {\rm eV}, g_{\phi \gamma}=2.5 \times 10^{-8} {\rm GeV^{-1}}$ and initial values of $\phi$ and $\dot{\phi}$ is equal to $\phi_i=10^{18} {\rm eV}, \dot{\phi_i}= 1 {\rm eV^2}$, respectively. Our results for this case are shown in \Fig{figure1}. In the upper panel of \Fig{figure1}, we show the evolution of the vector potential normalized to its initial value, $\mathcal{A}(z)/\mathcal{A}(0)$ for $k/k_c=0.2$ and $1.0$. The solid yellow color show the evolution of the `-' helicity mode and the dashed yellow color shows the evolution of the `+' helicity mode for $k/k_c=1.0$. The blue curve are for the case in which $k/k_c=0.2$. The lower panel of \Fig{figure1} demonstrate the evolution of the energy densities of $\phi$ (blue curve), $\rho_{\phi} = (m^2 \phi^2+\dot{\phi}^2)/2$ and EM field (yellow curve) normalized to the initial value of scalar field energy density. The EM field energy density is given by,
\begin{align}\label{mag_spec}
&\quad \quad\rho_{EM}=\rho_B+\rho_E.\\
\text{Here,}\nonumber\\
    \rho_B &\equiv \int d\ln k \frac{d \rho_B}{d \ln k}= \int d \ln k \frac{k^5}{(2\pi^2) }(|\mathcal{A}_+|^2+|\mathcal{A}_-|^2),\\
    \rho_E &\equiv \int d\ln k \frac{d \rho_E}{d \ln k}= \int d \ln k \frac{k^3}{(2\pi^2) }(|\dot{\mathcal{A}}_+|^2+|\dot{\mathcal{A}}_-|^2).
\end{align}
As evident from this figure, there is an amplification of the vector potential in each half cycle of the oscillating $\phi$ field. As the amplification continues with time, in a few cycles, the EM field energy density will be comparable to the $\phi$ field energy density, and the gauge field backreaction on the evolution of $\phi$ will be important. However studying the backreaction is not the aim of this letter (please see Ref.~\cite{Brahma:2026fre} for the backreaction effects). We choose our parameters similar to the parameters of Fig.2 of Ref.~\cite{Brahma:2026fre}. On comparing our results shown in figure 1 with Fig. 2 of \cite{Brahma:2026fre}, we can say that there is a good match between the two results.

\begin{figure}[h!]
\centering
 \includegraphics[width=0.5\textwidth]{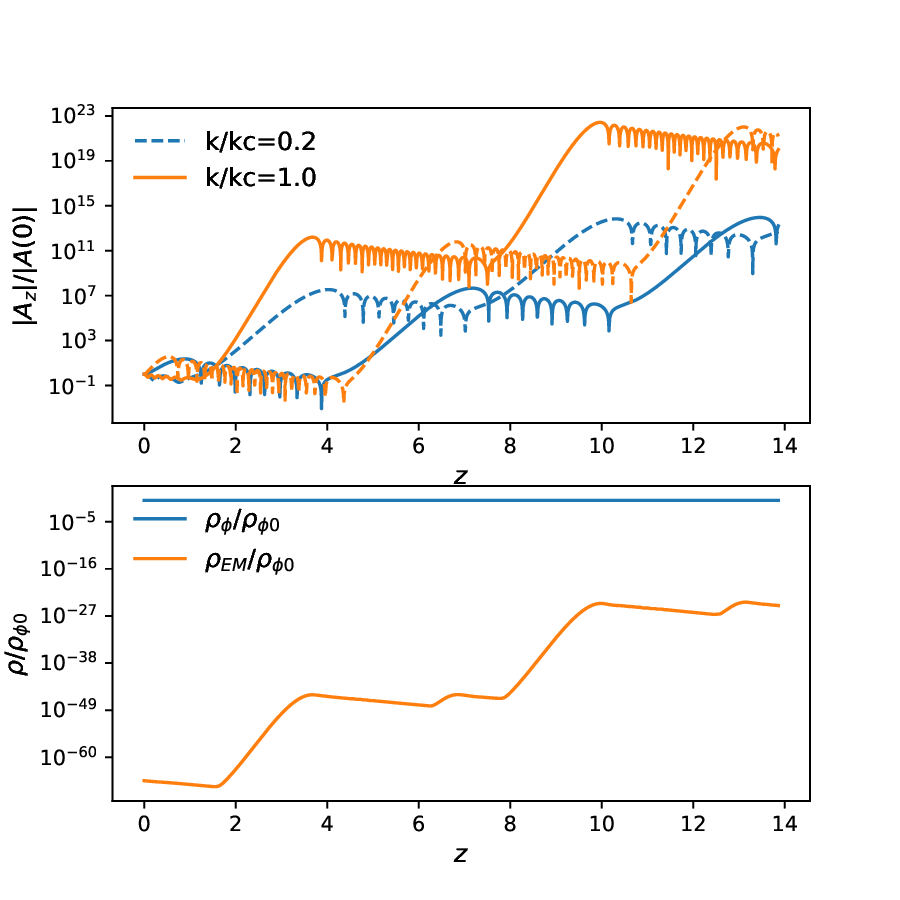}
\caption{Same as \Fig{figure1}, but for a non-zero conductivity. In this case, we choose $\sigma$ such that the initial ratio $\sigma/H = 10^{10}$.}
\label{figure2}
\end{figure}
\begin{figure}[h!]
\centering
 \includegraphics[width=0.45\textwidth]{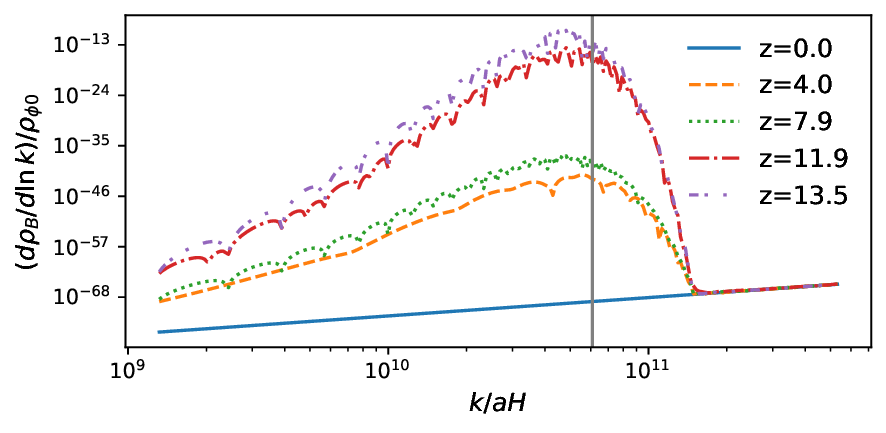}
 \includegraphics[width=0.45\textwidth]{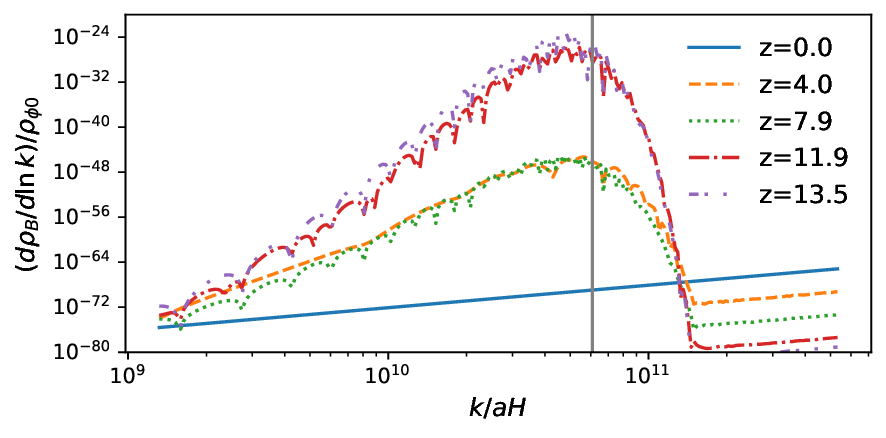}
  \includegraphics[width=0.45\textwidth]{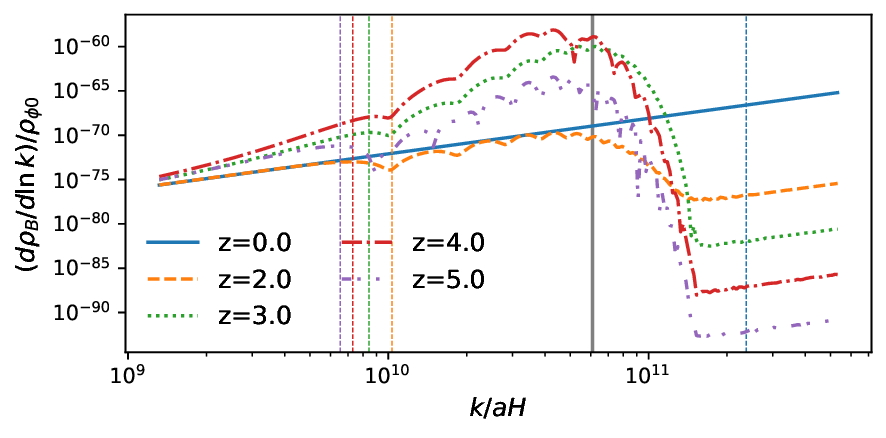}
\caption{The magnetic field spectrum is shown at different times, $z = 0,\, 4.0,\, 7.9,\, 11.9,$ and $13.5$. The vertical gray line indicates the wavenumber, $k/(aH) = g_{\phi\gamma} M_{\rm{pl}}$. The top panel corresponds to $\sigma = 0$, while the middle and lower panels is for initial values $\sigma/H = 10^{10}$ and $5 \times 10^{10}$, respectively.}
\label{figure3}
\end{figure}

In the second case, we include the plasma conductivity in the evolution of the EM field. The corresponding results are shown in \Fig{figure2}, where the panel layout and color scheme are the same as in \Fig{figure1}. For this case, we take the initial value $\sigma/H = 10^{10}$, which is much smaller than the value estimated in \Eq{sigma_by_H}, while keeping all other parameters identical to those used in \Fig{figure1}. As evident from the upper panel of \Fig{figure2}, the amplification of the vector potential is already significantly suppressed. Consequently, the energy density of the EM field remains much smaller than that of the $\phi$ field. These results are in agreement with the analytical expectations derived in \Eq{ampl_cond}.

Furthermore, we show the magnetic field spectrum defined in \Eq{mag_spec} in \Fig{figure3} at different times. To compute the spectrum, we solve \Eq{A_equation_with_plasma} for 256 values of the wavenumber around $k_c$. The top panel corresponds to the case $\sigma = 0$, while the middle and lower panels show results for initial values $\sigma/H = 10^{10}$ and $5 \times 10^{10}$, respectively. The vertical gray line marks the scale defined by $k/(aH) = g_{\phi\gamma} M_{\mathrm{pl}}$, i.e., $k = k_c$. From the top panel, we observe that the spectrum peaks around $k \sim k_c$, consistent with the discussion in the previous section, and exhibits a wavenumber-dependent amplification for modes with $k \lesssim k_c$. The inclusion of conductivity introduces two effects. First, it suppresses the amplification of the EM field arising from parametric resonance, as already discussed in the simplified analysis of the previous section. Second, it leads to damping of modes with $k \gtrsim k_d \sim \sqrt{\sigma H}$ due to dissipative effects in the plasma. In the lower panel, the dashed vertical line corresponds to the damping scale, $k/(aH) \sim \sqrt{\sigma/(t H^2)}$ for the corresponding value of $t = z/m$.

As evident from the bottom panel of \Fig{figure3}, the spectrum begins to decrease after $z \simeq 4$. This indicates that the amplification can no longer compete with dissipative damping. This behavior is observed for $\sigma/H \sim 5 \times 10^{10}$. For the more realistic value $\sigma/H \sim 10^{26}$ given in \Eq{sigma_by_H}, the amplification is expected to be significantly weaker, since the growth exponent $\mu_k$ is suppressed by the factor $H/\sigma$, as shown in \Eq{ampl_cond}. Although the damping scale shifts to much larger wavenumbers and the spectral peak remains well below this scale, the overall amplification is extremely small. We therefore conclude that, for realistic values of $\sigma/H$, modes around $k \sim k_c$ do not undergo significant dissipative damping; however, their amplification is also negligible. Consequently, this scenario does not lead to magnetogenesis of astrophysical relevance.

\noindent \textbf{Conclusion.---}
In this letter, we study the post-recombination dynamics of an ultralight pseudoscalar field coupled to the EM field via the term $g_{\phi \gamma} \phi F_{\mu \nu}\tilde{F}^{\mu\nu}$ in the Lagrangian density, to explore the possibility of magnetogenesis via parametric resonance driven by the oscillating scalar field. In recent studies, Ref.~\cite{Brandenberger:2025gks,Brahma:2026fre}, the authors found that this mechanism can efficiently generate magnetic fields whose energy density constitutes a significant fraction of the dark matter energy density for an observationally viable coupling strength of the pseudoscalar and EM field. However, the analysis in Ref.~\cite{Brandenberger:2025gks} neglects the effects of the conducting medium present after recombination.

In this work, we incorporate the finite conductivity of the plasma and find that the amplification of the EM is significantly suppressed, scaling as the ratio of the Hubble parameter to the conductivity. This ratio is extremely small, of order $10^{-26}$ at a temperature $T \sim 0.3~{\rm eV}$. By requiring the growth rate to exceed the Hubble expansion rate, we obtain the condition $g_{\phi \gamma} \gtrsim 10^{-6}\,{\rm GeV^{-1}}$ for $H/\sigma \sim 10^{-25}$, a typical value of the conductivity in the post-recombination epoch. However, such a large coupling is incompatible with existing observational constraints on $g_{\phi\gamma}$ for ultralight pseudoscalar dark matter coupled to the EM field \cite{Reynolds:2019uqt,Reynes:2021bpe,CAST:2024eil,Fedderke:2019ajk}. We therefore conclude that parametric resonance driven by the oscillating dark matter field does not result in significant amplification, rendering this mechanism ineffective for generating magnetic fields of astrophysical interest.

{
	\textbf{Acknowledgments.---}
}
\label{sec:acknowledgments}
RS would like to thank the Department of Science and Technology, India, for financial support through the INSPIRE Faculty Fellowship. The authors are grateful to Kandaswamy Subramanian for helpful discussions on post-recombination conductivity and for guiding us to the relevant literature.
\bibliographystyle{apsrev4-2} 
\bibliography{reference}

\end{document}